\documentclass[twocolumn,english,showpacs,preprintnumbers]{revtex4}

\input{epsf}
\usepackage{graphicx}

\def\AJ{{\it Astroph. J.} }

\def\GRG{{\it Gen. Relativity and Gravitation} }

\def\PL{{\it Phys. Lett.} }
\def\PR{{\it Phys. Rev.} }

\def\frac#1#2{{\textstyle{{#1}\over {#2}}}}

\def\lsim{\mathrel{\rlap{\lower4pt\hbox{\hskip1pt$\sim$}}
    \raise1pt\hbox{$<$}}}
\def\gsim{\mathrel{\rlap{\lower4pt\hbox{\hskip1pt$\sim$}}
    \raise1pt\hbox{$>$}}}
\def\sqr#1#2{{\vcenter{\vbox{\hrule height.#2pt
         \hbox{\vrule width.#2pt height#1pt \kern#1pt
         \vrule width.#2pt}
         \hrule height.#2pt}}}}

\def\beq{\begin{equation}}
\def\eeq{\end{equation}}
\def\beqa{\begin{eqnarray}} 
\def\eeqa{\end{eqnarray}}

\def\laq{\raise 0.4 ex \hbox{$<$}\kern -0.8 em\lower 0.62 ex\hbox{$\sim$}}
\def\gaq{\raise 0.4 ex \hbox{$>$}\kern -0.7 em\lower 0.62 ex\hbox{$\sim$}}
\makeatother
\begin{document}

\preprint{DF/IST-4.2004}

\title{Challenges to the Generalized Chaplygin Gas Cosmology }

\author{O. Bertolami}

\affiliation{Departamento de F\'\i sica, Instituto Superior T\'ecnico\\
Av. Rovisco Pais 1, 1049-001 Lisboa, Portugal}

\date{\today}


\begin{abstract}
The generalized Chaplygin gas (GCG) model allows for an unified description of 
the cosmologically recent accelerated expansion of the Universe and of the evolution of energy 
density 
perturbations. This dark energy - dark matter unification is achieved through 
a rather exotic background fluid whose equation of state is given by
$p = - A/\rho^{\alpha}$, where $A$ is a positive constant.
Observational constraints arising from bounds on 
the locations of the first few peaks and troughs of the Cosmic Microwave
Background Radiation (CMBR) power spectrum from recent WMAP and
BOOMERanG experiments are consistent with the model for $\alpha \lsim 0.6$ assuming that 
$0 < \alpha \le 1$. Most recent Type-Ia Supernova data indicates however, that the range $\alpha > 1 $ must be considered.
\end{abstract}
\maketitle

\section{Introduction}

In this brief review I shall summarize some recent results obtained in the context of the GCG model 
and discuss some challenges this dark energy - dark matter unification model face. 
The results presented here largely rely on the work developed in Refs. \cite{Bento1,Bento2,Bento3,Bento4,Silva,Bertolami}.

Cosmology is undergoing a particularly fruitful period. Recent precision measurements obtained 
by dedicated experiments on the Cosmic Microwave background Radiation (CMBR), supernova searches and 
large galaxy surveys allow for detailed comparisons with the theoretical 
models. It is remarkable that all available data can be accounted by the {\it Hot Big Bang Model} enriched 
with {\it Inflation}, a period of accelerated expansion in the very early 
Universe that reconciles cosmology with causality and that elegantly explains the origin of the observed 
Large Scale Structure of the Universe. However, in order to fully understand the observations, 
at least two additional new entities are required: {\it Dark Matter} and {\it Dark Energy}. Dark matter was 
originally proposed to explain the rotation curves of galaxies and it turns out 
to be a fundamental building block for structure formation at large 
scales. Dark energy corresponds to a smoothly distributed energy that cannot be  
related with any known form of matter and is required  
to explain the recently observed accelerated expansion of the Universe. 
Even though presumably the inflationary process has been caused by 
the the dynamics of a scalar field, the inflaton, 
and that the underlying structures behind 
dark energy and dark matter might also be scalar fields,  
these three concepts are apparently unrelated. However, in what concerns dark matter and dark energy, 
a scheme has emerged where an unification of these entities is 
possible through a perfect fluid description with an exotic equation of state:

\beq
p_{ch} = - {A \over \rho_{ch}^\alpha}~~,
\label{eq:eqstate}
\eeq
\vskip 0.3cm

\noindent
where $A$ and $\alpha$ are positive constants. Most of the work on GCG cosmology has assumed that 
$0 < \alpha \le 1$, however, more recently, in studying the latest Type-Ia Supernova data 
it has emerged that $\alpha > 1$ values are 
preferred. This equation of state with $\alpha=1$ was first introduced  
in 1904 by the Russian physicist Chaplygin to describe
aerodynamic processes \cite{Chaplygin}; its importance to cosmology was pointed out 
in Ref. \cite{Kamenshchik}. The suggestion of its generalization for $\alpha\neq 1$ was also
proposed in Ref. \cite{Kamenshchik} and the ensuing  cosmology has been analyzed in
Ref. \cite{Bento1}. It is remarkable that the Chaplygin equation of state has a well defined connection with string 
and brane theories (see Ref. \cite{Jackiw} for a through review). 

The idea that a cosmological model based on the Chaplygin gas could
lead to the  unification of dark energy 
and dark matter, was first advanced for the $\alpha=1$ case in Refs. \cite{Bilic,Fabris}, 
and generalized to $\alpha \neq 1$ in Ref. \cite{Bento1}.

\section{The Model}

The reason why the cosmological features of the equation of state (\ref{eq:eqstate}) 
are so interesting can be 
better appreciated after inserting it 
into the relativistic energy-momentum conservation equation, which yields for the 
evolution of the energy density \cite{Bento1}

\beq
\rho_{ch} =  \left[A + {B \over a^{3 (1 + \alpha)}}\right]^{1 \over 1 +
 \alpha}~~,
\label{eq:rhoc}
\eeq 
\vskip 0.3cm

\noindent
where $a$ is the scale-factor of the Universe and $B$ an integration 
constant. It is easy to see that Eq.~(\ref{eq:rhoc}) is directly related with the observed
accelerated expansion of the Universe as it automatically 
leads to an asymptotic phase where the equation of state is dominated by a 
cosmological constant, $8 \pi G A^{1/1+\alpha}$, while at earlier times 
the energy density behaves as if dominated by non-relativistic
matter. This dual 
behaviour is at the heart of the unification scheme provided by the GCG model. 
This unification can be achieved through an underlying description based on a complex scalar field model 
which admits an inhomogeneous generalization. This generalization is shown to be consistent 
with standard structure formation scenarios \cite{Bento1,Bilic,Fabris}. 
It is clear that the GCG model corresponds to the $\Lambda$CDM model 
for $\alpha = 0$ (and also $A_s = 1$; see Eq. (\ref{Hubblexp}) below).

These remarkable properties make the 
GCG model an interesting alternative to models where the 
accelerated expansion of the Universe arises from  
an uncanceled cosmological constant or a rolling scalar field as in quintessence
models.

From Eq.~(\ref{eq:eqstate}) one can see that, in principle, any 
positive $\alpha$ values are admissible. The range $0 < \alpha \le 1$, has been chosen so that 
the sound velocity ($c_s^2 = \alpha A/ \rho_{ch}^{1+\alpha}$) does not exceed the velocity of light,
in the regime where the effective equation of state has the form of ``soft'' matter, $p = \alpha \rho$, 
in which case $c_s^2 = \alpha$. Notice however, that as described, in 
Ref. \cite{Bertolami}, one can accommodate the case 
$\alpha > 1$ in a manifestly Lorentz invariant underlying theory that does not 
violate the dominant energy condition $\rho + p \ge 0$.

As already mentioned, at a more fundamental level, the GCG model can be described 
by a complex scalar field whose Lagrangian density can be written in the form of a generalized 
Born-Infeld Lagrangian density. This can be seen starting with the Lagrangian density for a
massive complex scalar field, $\Phi$,
 
\beq
{\cal L} = g^{\mu \nu} \Phi^{*}_{, \mu} \Phi_{, \nu} - V(\vert \Phi \vert^2)~~,
\label{eq:complexfield}
\eeq
\vskip 0.3cm
\noindent
expressed in terms of its mass, $M$, 
as $\Phi = (\phi / \sqrt{2}m) \exp(- iM \theta)$. Considering
the scale of the inhomogeneities as corresponding to the spacetime variations of
$\phi$ on scales greater than $M^{-1}$, then $\phi_{, \mu} << M \phi$,
which, together with Eq.(\ref{eq:eqstate}), yields to the relationship:

\beq
\phi^2(\rho_{ch}) = \rho_{ch}^{\alpha} 
(\rho_{ch}^{1 + \alpha} - A)^{{1 - \alpha \over 1 + \alpha}}~~,
\label{eq:phidens}
\eeq
\vskip 0.3cm
\noindent
following that the Lagrangian density Eq. (\ref{eq:complexfield}) assumes the form of a 
{\it generalized} Born-Infeld Lagrangian density:

\beq
{\cal L}_{GBI} = - A^{1 \over 1 + \alpha} 
\left[1 - (g^{\mu \nu} \theta_{, \mu} 
\theta_{, \nu})^{1 + \alpha \over 2\alpha}\right]^{\alpha \over 1 + \alpha}~~.
\label{GenBorn-Infeld} 
\eeq
Notice that, for $\alpha=1$, one recovers the exact Born-Infeld Lagrangian density.

Alternatively, the GCG model can be described by a minimally coupled scalar field, $\varphi$, 
with canonical kinetic energy term and a potential of the form \cite{Bertolami}:

\beq
V = V_0e^{3(\alpha - 1)} \left[cosh({m \varphi \over 2})^{2/(\alpha + 1)} 
+ cosh({m \varphi \over 2})^{-2/(\alpha + 1)}\right]~~,
\label{Potential} 
\eeq
where $V_0$ is a constant and $m=3(\alpha + 1)$. In this case, 
$c^2_s = 1$, irrespective of the value of $\alpha$.

In what follows, we shall discuss the observational bounds that can be set on 
the GCG  model parameters.

\section{Observational Constraints}

Given that the GCG model stands out as a potentially viable dark energy - dark
matter unification scheme many authors have developed methods aiming to  
constrain its parameters from observational data,   
particularly through SNe Ia data \cite{Supern}, CMBR peak and through location \cite{Bento2,Bento3} 
and amplitudes \cite{Finelli},
and gravitational lensing statistics \cite{Silva,Alcaniz}. More recent analysis 
based on the latest Type-Ia Supernova data has yielded rather surprising results, namely that 
$\alpha > 1$. 

Particularly stringent constraints arise from the study of the position
of the acoustic peaks and troughs of the CMBR power spectrum. The CMBR
peaks arise from oscillations of the primeval plasma just before the
Universe becomes transparent. Driving processes and the ensuing shifts
on peak positions can be written as \cite{Hu}

\beq 
\ell_{p_m} \equiv \ell_A \left(m - \varphi_m\right)~, 
\label{eq:lm}
\eeq
where $\ell_A$ is the acoustic scale

\beq
\label{eq:la}
\ell_A = \pi {\tau_0 - \tau_{\rm ls} \over \bar c_s \tau_{\rm ls}}~~,
\eeq 
and with $\tau_0$ and $\tau_{\rm ls}$ standing for the conformal time ($\tau =
\int a^{-1} dt$) today and at last scattering and $\bar{c}_s$ the
average sound speed before decoupling. Given that peak shift processes are fairly independent of 
physics after recombination they are not affected by the  
nature of the late time acceleration mechanism. Thus, the 
accurate fitting formulas of Ref.~\cite{Doran} can be used to compute
the phase shifts $\varphi_m$ for the GCG model. In order to estimate
the acoustic scale, we use Eq. (\ref{eq:rhoc}) and write the Universe
rate of expansion as
 
\beq
\label{eq:H2}
H^2={8\pi G\over 3}\left[{\rho_{r0}\over a^4}+{\rho_{b0}\over{a^3}}+\rho_{ch0}
\left( A_s + 
{(1-A_s)\over a^{3(1+\alpha)}}\right)^{1/1+\alpha}\right]~~,
\label{Hubblexp}
\eeq
\vskip 0.3cm
\noindent
where $A_s\equiv A/ \rho_{ch0}^{1+\alpha}$, $\rho_{ch0}\equiv
(A+B)^{1/ 1+\alpha}$. We have included the contribution of
radiation and baryons as these are not accounted for by the GCG
equation of state. As discussed in Refs. \cite{Bento2,Bento3}, the
above equations allow for obtaining the value of the
fundamental acoustic scale by direct integration, using that
$H^2=a^{-4} \left(d a\over d \tau\right)^2$.

Comparison of the outcome from the above procedure with the most recent results on the 
location of the first two peaks and the first trough obtained by the 
WMAP collaboration \cite{WMAP}, namely 
$\ell_{p_1} = 220.1\pm 0.8,~ \ell_{p_2} = 546\pm 10,~ \ell_{d_1} = 411.7\pm 3.5$, with 
the bound on the position of the third peak obtained by the 
BOOMERanG collaboration \cite{Boomerang}, $l_{p3}=825^{+10}_{-13}$, gives origin 
to quite strong constraints on the model parameters. These constraints can be summarized as follows and 
critically depend on values of the spectral tilt, $n_s$ and of the Hubble parameter, $h$ \cite{Bento3,Bento4}:

\vskip 0.3cm
\noindent
1) Assuming WMAP priors, the Chaplygin gas model, $\alpha = 1$, is incompatible
with the data and so are models with $\alpha \gsim 0.6$.

\vskip 0.3cm
\noindent
2) For $\alpha = 0.6$, consistency 
with data requires for the spectral tilt, $n_s > 0.97$, and that, $h \lsim 0.68$. 

\vskip 0.3cm
\noindent
3) The $\Lambda$CDM model barely fits the data for values of the spectral tilt 
$n_s \simeq 1$ (WMAP data yields $n_s=0.99\pm0.04$) and for that $h > 0.72$. 
For low values of $n_s$, $\Lambda$CDM is preferred to the GCG models, 
whereas for intermediate values of $n_s$, the GCG model is favoured only if $\alpha \simeq 0.2$. 

\vskip 0.3cm
\noindent
These results are essentially consistent with the ones obtained in Refs. \cite{Finelli} using the CMBFast code.
Furthermore, we find that:

\vskip 0.3cm
\noindent
4) Our study of the peak locations in the $(A_s,\alpha)$ plane shows
that, varying $h$ within the bounds $h=0.71^{+0.04}_{-0.03}$
\cite{WMAP}, does not lead to very relevant changes in the allowed
regions, as compared to the value h=0.71, even though
these regions become slightly larger as they shift up-wards for
$h<0.71$; the opposite trend is found for $h>0.71$.

\vskip 0.3cm
\noindent
5) Our results are consistent with bounds obtained in
Ref. \cite{Bento2} using BOOMERanG data for the third peak and Archeops
\cite{Benoit} data for the first peak as well as results from SNe Ia \cite{Supern}
and age bounds, namely $0.81 \lsim A_s \lsim 0.85$ and $0.2 \lsim
\alpha \lsim 0.6$.

\vskip 0.3cm
\noindent
6) If one abandons the constraint on $h$ arising from WMAP, then 
the Chaplygin gas case $\alpha = 1$ is consistent with the peaks location, if $h \le 0.64$. 
\cite{Bento3}.

\vskip 0.3cm
\noindent
Quite challenging, a new set of constraints arise from the latest SNe Ia data. These arise from the study 194 
supernova data points from Ref. \cite{Tonry}. The results can be summarized as follows \cite{Bertolami}:

\vskip 0.3cm
\noindent
7) Data favours $\alpha > 1$, although there is a strong degeneracy on $\alpha$. At $68 \%$ confidence level the 
minimal allowed values for $\alpha$ and $A_s$ are $0.78$ and $0.778$, thus ruling out the $\Lambda$CDM model 
$\alpha = 0$ case. However, at $95 \%$ confidence level there is no constraint on $\alpha$.

\vskip 0.3cm
\noindent
8) If one does not assume the flat prior, one finds that GCG is consistent with data for values of $\alpha$ sufficiently 
different from zero. Allowing some small curvature, positive or negative, one finds that the GCG model is a more suitable 
description than the $\Lambda$CDM model.

\vskip 0.3cm
\noindent
These results are analogous to the ones obtained in Refs. \cite{Sahni,Choudhury} which find that the latest supernova 
data favours ``phantom''-like matter with an equation of state of the form $p = \omega \rho$ with $\omega < -1$.

\section{Conclusions and Outlook}

In this brief review we have outlined the way the GCG model allows for a
consistent description of the accelerated expansion of the Universe and 
suggests an interesting and promising scheme for the unification of dark energy and dark
matter. The model is quite detailed and its predictions can be directly confronted 
with observational data.  For this purpose,
several studies were performed aiming to constrain the parameter space
of the model using Supernovae data, the age of distant quasar sources,
gravitational lensing statistics and the location of the first few
peaks and troughs the CMBR power spectrum, as measured by the WMAP and
BOOMERanG collaborations. These studies reveal that a substantial portion
of the parameter space of the GCG model can be excluded. In these studies it has been assumed that 
$0 < \alpha \le 1$, however, a recent study of the latest supernova data indicates that $\alpha > 1$ values 
are favoured. One can see that there is no contradiction between the various observational constraints at $2 \sigma$ 
level, even though a full analysis is still missing.
 
A critical question for the GCG model concerns structure formation. This is at the heart of the model as it is 
meaningful only as an entangled mixture of dark matter and dark energy. 
Concerns about this issue have been raised 
\cite{Sandvik}, however in this analysis the effect of baryons has not been taken into account, 
which was shown to be relevant and necessary for consistency with the 2DF mass
power spectrum \cite{Beca}. Furthermore, most computations were based on the
linear treatment of perturbations close to the present time, thus neglecting
any non-linear effects which are clearly important. Moreover, the role of entropy perturbations \cite{Reis} in the 
non-linear regime and the comparison with observable quantities has to be further examined \cite{Waga}.

\vskip 1cm
\begin{acknowledgments}

\vskip 0.2cm

\noindent
It is a pleasure to thank my collaborators Maria da Concei\c c\~ao Bento, Anjan Sen, Somasri Sen and 
Pedro Tavares Silva for making the work on the GCG model so enjoyable. 
I would like also thank Ja\'\i lson Alcaniz, J\'ulio Fabris, Martin 
Makler and Ioav Waga for explaining me their interesting and challenging results on various aspects of 
the GCG model. I am in great debt to 
Tom Girard for having invited me to talk at the Portuguese
COSLAB Workshop in Oporto on May 2003, Ana Ach\'ucarro for the 
invitation to talk at the Third COSLAB Workshop, First Joint COSLAB - VORTEX - BEC2000+ 
Workshop in Bilbao on July 2003 and Reuven Opher for the invitation to attend 
and talk at the III Workshop: Nova F\'\i sica no Espa\c co, Campos do Jord\~ao, Brazil, February 2004.
I have to thank the Funda\c c\~ao para a 
Ci\^encia e a Tecnologia (Portugal) for the partial support 
under the grant POCTI/FIS/36285/2000. 

\end{acknowledgments}

\vskip 1cm

\end{document}